\documentclass[conference]{IEEEtran}
\IEEEoverridecommandlockouts
\usepackage{cite}
\usepackage{amsmath,amssymb,amsfonts}
\usepackage{algorithmic}
\usepackage{graphicx}
\usepackage{textcomp}
\usepackage{xcolor}
\usepackage{fitbox}
\usepackage{subcaption}   

\usepackage{tcolorbox}
\usepackage{pbox}

\def\BibTeX{{\rm B\kern-.05em{\sc i\kern-.025em b}\kern-.08em
    T\kern-.1667em\lower.7ex\hbox{E}\kern-.125emX}}
\begin{document}

\makeatletter
\newcommand{\linebreakand}{%
  \end{@IEEEauthorhalign}
  \hfill\mbox{}\par
  \mbox{}\hfill\begin{@IEEEauthorhalign}
}
\makeatother

\title{Understanding the Structure of QM7b and QM9 Quantum Mechanical Datasets Using Unsupervised Learning\\
}


 \author{
 \IEEEauthorblockN{Julio J. Valdés}
 \IEEEauthorblockA{\textit{National Research Council Canada}  \\
 \textit{Digital Technologies Research Centre} \\
 Ottawa, Canada \\
 julio.valdes@nrc-cnrc.gc.ca}
 \and
 \IEEEauthorblockN{Alain B. Tchagang}
\IEEEauthorblockA{\textit{National Research Council Canada}  \\
 \textit{Digital Technologies Research Centre} \\
 Ottawa, Canada \\
 alain.tchagang@nrc-cnrc.gc.ca}
}

\maketitle

\begin{abstract}
This paper explores the internal structure of two
quantum mechanics datasets (QM7b, QM9), composed of several thousands of organic molecules and described in terms of electronic properties. Understanding the structure and characteristics of this kind of data is important when predicting the atomic composition from the properties in inverse molecular designs.
Intrinsic dimension analysis, clustering, and outlier detection methods were used in the study. They revealed that for both datasets the intrinsic dimensionality is several times smaller than the descriptive dimensions. The QM7b data is composed of well defined clusters related to atomic composition. The QM9 data consists of an outer region predominantly composed of outliers, and an inner core region that concentrates clustered, inliner
objects. A significant relationship exists between the number of atoms in the molecule and its outlier/inner nature. Despite the structural differences, the predictability of variables of interest for inverse molecular design is high. This is exemplified with models estimating the number of atoms of the molecule from both the original properties, and from lower dimensional embedding
spaces.

\end{abstract}


\section{Introduction}
\label{Introduction}

The goal of molecular design (MD) is to generate molecules with desired properties, which would correspond to new materials, new drugs, and compounds of interest. This is a very complex undertaking considering the size of the chemical compound space, where conventional approaches are either inefficient, time-consuming, and ultimately prohibitive. In that sense, the introduction of technologies leveraging the progress made in computationally intensive data analysis and artificial intelligence disciplines (e.g. machine learning and others) represents an important step towards achieving the aforementioned goal.

Specifically from a machine learning perspective, the two main approaches are the so-called discriminative and generative. In the first case (also known as forward design, the purpose is to relate molecules with their properties (electronic or otherwise), whereas, in the generative approach (inverse design), the idea is to start with a given collection of properties and derive the actual molecules compatible with them \cite{Rupp-2015-Machine-learning-for-quantum-mechanics-in-a-nutshell}. 
The latter has been approached with different techniques, including deep learning and others. 

Variational autoencoders (VAE) working with real-valued vectors representing the molecules have been used for property optimization in latent spaces \cite{Gomez-Bombarelli-et-al-2018-Automatic-chemical-design-using-a-data-driven-continuous-representation-of-molecules}. The targeted application was the generation of new molecules for efficient exploration and optimization through open-ended spaces of chemical compounds with molecules having fewer that nine heavy atoms.
A comparison of a VAE with an adversarial autoencoder (AAE) was presented in \cite{Blaschke-et-al-2017-Application-of-generative-autoencoder-in-de-novo-molecular-design} and applied to generate novel compounds with activity against dopamine receptor type 2.
VAE and AAE were also used in \cite{Kadurin-et-al-2017-druGAN} with the purpose of identifying new molecular fingerprints with predefined anticancer properties, showing that the models significantly enhanced the efficiency of the development of new molecules with specific anticancer properties.
The application of other approaches to inverse molecular design such as generative adversarial networks (GAN), reinforcement learning  and transfer learning have been described in \cite{Jorgensen-et-al-2018-Deep-Generative-Models-for-Molecular-Science}.
A classification formulation of the inverse molecule design problem following the direct approach was investigated in \cite{Tchagang-and-Valdes-2022-Towards-a-Classification-Scheme-for-Inferring-the-Atomic-Composition-of-Drug-like-Molecules}, where high-quality models for predicting atomic composition from electronic properties were found.
However, there is a lack of studies oriented to the understanding of the structure of the molecule datasets used in the aforementioned approaches. Gaining insights related to the internal structure of the information and the characteristics of the data could benefit inverse molecule design and related topics. The objective of this paper is to start filling that gap by using unsupervised machine learning methods on two of the datasets (QM7b and QM9) that relate quantum mechanical properties with atomic composition and that have been used in inverse molecular design studies.


The rest of this paper is organized as follows: Section~\ref{Molecule Datasets} describes the data used in the study, Section~\ref{Machine Learning Methods} describes the collection of machine learning methods applied, Section~\ref{Results} presents the results and Section~\ref{Conclusions} the conclusions.

\section{Molecule Datasets}
\label{Molecule Datasets}
QM7b and QM9 are the two molecule datasets used in the paper. They are part of a larger collection of resources available in \cite{quantum-machine_org-datasets} with the aim of accelerating the development of fast and accurate simulations of quantum-chemical systems from first principles.

The QM7b dataset  \cite{Blum-and-Reymond-2009-QM7b}, \cite{Montavon-et-al-2013-QM7b} is a derivation of the QM7 data, which is  composed of all molecules of up to 23 atoms (including 7 heavy atoms C, N, O, and S), totalling 7165 molecules. This dataset features a large variety of molecular structures such as double and triple bonds, cycles, carboxy, cyanide, amide, alcohol and epoxy. QM7b is an extension of QM7 oriented to for multitask learning where 13 additional properties (e.g. polarizability, HOMO and LUMO eigenvalues, excitation energies) have to be predicted at different levels of theory (ZINDO, SCS, PBE0, GW). Additional molecules comprising chlorine atoms are also included, totaling 7211 molecules.
QM7b is composed of 7211 objects described in terms of 14 properties representing the inputs (Coulomb matrices). Each molecule is composed of atoms from the following six chemical elements: Carbon (C), Chlorine (Cl), Hydrogen (H), Nitrogen (N), and Oxygen (O), and Sulfur (S). The following quantum-derived properties describe the molecules: 1) PBE0 atomization energies, 2) zindo excitation energy, 3) zindo highest absorption intensity, 4) zindo homo, 5) zindo lumo, 6) zindo 1st excitation energy, 7) zindo ionization potential, 8) zindo electron affinity, 9) PBE0 homo, 10) PBE0 lumo, 11) GW homo, 12) GW lumo, 13) PBE0 polarizability A3, 14) SCS polarizability A3.


The other dataset used in this study is known as the QM9 dataset \cite{ruddigkeit-2012-Enumeration-of-166-billion-organic-small-molecules}, \cite{ramakrishnan-et-al-2014-Quantum-chemistry-structures-and-properties-of-134-kilo-molecules}. It contains $N=133,885$ small organic molecules, where each molecule is composed of a combination of the following five chemical elements: Carbon (C), Chlorine (Cl), Hydrogen (H), Nitrogen (N), and Oxygen (O). In inverse design approaches, they represent model targets. The molecules are characterized by $N_v = 19$ electronic properties (geometric, energetic, electronic, and thermodynamic), computed using quantum chemistry techniques. Contrary to  QM7b which has none, QM9 has 6,095 constitutional isomers among the 134k molecules.
It is considered that this dataset provides very accurate quantum chemical properties for a relevant, consistent, and comprehensive chemical space of small organic molecules. The properties are
given by: 1) Rotational Constant, 2) Rotational Constant, 3) Rotational Constant, 4) Norm of dipole moment, 5) Norm of static polarizability, 6) Highest unoccupied molecular orbital, 7) Lowest unoccupied molecular orbital,8) Difference between homo and lumo, 9) Electronic spatial extent, 10) Zero point vibrational energy, 11) Internal energy at 0K, 12) Internal energy at 298.15K, 13) Enthalpy at 298.15k, 14) Free energy at 298.15K, 15) Heat Capacity, 16) Atomization energy at 0K, 17) Atomization energy at 298.15K, 18) Atomization enthalpy at 298.15k, 19) Free atomization energy at 298.15K

QM9 has been used for a variety of tasks, like benchmarking existing methods, hybrid quantum mechanics/machine learning, and the identification of structure-property relationships. In this paper, the focus is on characterizing the internal structure of the data from a machine-learning perspective and on assessments for inferring the atomic composition from molecular properties.

\subsection{Preprocessing}
\label{Preprocessing}
For both QM7b and QM9, the original properties describing the data are expressed in different units of measure. Variables that are measured at different scales do not contribute equally to the analysis and create biases, affecting a broad variety of statistical and machine-learning methods. In order to make their values comparable, and also to eliminate this source of bias, prior to the application of the machine learning methods described in the following section, each dataset was preprocessed by converting the original property values to z-scores. With the new variables having zero mean and unit variance, all properties are expressed in the same unit of measure (their variance), eliminating that source of bias.

\section{Machine Learning Methods}
\label{Machine Learning Methods}

The machine learning methods used in the study are composed of unsupervised and supervised techniques. The former were aiming at creating alternative data representations and their characterization, including the construction of visual representations and nonlinear mappings for manifold extraction. The latter consisted of prediction algorithms applied to the different data representations.

\subsection{Unsupervised Learning}
\label{mlmethods:Unsupervised}
\subsubsection{Intrinsic Dimension}
\label{mlmethods:Intrinsic-Dimension}

Data is typically represented in two forms: as objects described by their features, or as a collection of pairwise similarities or dissimilarities. In real-world problems, the dimensionality of these spaces can be large, leading to the curse of dimensionality, which can negatively impact the performance of statistical and machine learning algorithms. However, the information contained in the data is often concentrated in a subspace of much lower dimensionality than that of the original descriptor space. The intrinsic dimension (IDim) is the dimension of an embedded manifold that contains the data. Identifying these manifolds can aid in understanding the internal structure of the information, and working with them can alleviate the curse of dimensionality and improve the performance of machine learning methods. 

Finding the intrinsic dimension has no univocal solution. Estimates based on different approaches vary and it is necessary to work with a consensus from the estimated values, since they are based on a variety of principles and assumptions. The following approaches were used in this paper for estimating the intrinsic dimension of the QM9 dataset: 
%
%
Angle and norm concentration (DANCo) \cite{ceruti-2014-DANCo_An-intrinsic-dimensionality-estimator-exploiting-angle-and-norm-concentration}, Correlation Integral (CorrIn) \cite{grassberger-1983-Measuring-the-strangeness-of-strange-attractors}, Fisher separability (FisherS) \cite{albergante-et-al-2019-Estimating-the-effective-dimension-of-large-biological-datasets-using-Fisher-separability-analysis}, \cite{gorban-et-al-2018-Correction-of-AI-systems-by-linear-discriminants_Probabilistic-foundations}, Minimum neighbor distance-ML (MiND\_ML) \cite{rozza-et-al-2012-Novel-high-intrinsic-dimensionality-estimators}, Maximum likelihood (MLE) \cite{levina-et-al-2004-Maximum-Likelihood-estimation-of-intrinsic-dimension}, Estimation within tight localities (TLE) \cite{amsaleg-et-al-2019-Intrinsic-dimensionality-estimation-within-tight-localities}, Method of Moments (MOM) \cite{amsaleg-et-al-2018-Extreme-value-theoretic-estimation-of-local-intrinsic-dimensionality}, PCA Fukunaga-Olsen (lPCA) \cite{fukunaga-and-olsen-1971-An-algorithm-for-finding-intrinsic-dimensionality-of-data}, \cite{fan-et-al-2010-Intrinsic-dimension-estimation-of-data-by-principal-component-analysis}, Minimal Neighborhood Information (TwoNN) \cite{facco-et-al-2017-Estimating-the-intrinsic-dimension-of-datasets-by-a-minimal-neighborhood-information}. They are implemented in \cite{bac-et-al-2021-Scikit-Dimension_A-Python-Package-for-Intrinsic-Dimension-Estimation} under a common framework.

The estimates of the intrinsic dimension provide an idea of the required dimension of embedding spaces containing the data. When they are based only on the information contained in the predictor features,  the new sets of generated features can be used as predictors for modeling, when the target variables are added.

\subsubsection{Manifold Learning}
\label{mlmethods:Manifold-Learning}
A low-dimensional transformation of the original representation space provides an environment where the information could be visualized and/or modeled with supervised methods. When the objective is to visualize the data, if the intrinsic dimension of the data is sufficiently low, the new space could provide useful insights into the structure of the information. Otherwise, it would provide an approximation. 
Whenever estimates of the intrinsic dimension are available, they could be used for making appropriate choices for the dimension of the embedding space via nonlinear transformations. Principles used for computing these transformations are, for example, the preservation of local distances/dissimilarities, the preservation of conditional probability distributions within neighborhoods, or more complex criteria. 
This paper used the Uniform Manifold Approximation and Projection (UMAP) which is a general-purpose manifold learning and dimension reduction algorithm \cite{mcinnes2018arXivUMAP,mcinnes2018umap-software}. It can be used for nonlinear mapping, non-linear dimension reduction, and visualization purposes. This state-of-the-art algorithm is based on a few assumptions: \textit{i)} The data is uniformly distributed on a Riemannian manifold (one with a collection of inner products for every point on a tangent space); \textit{ii)} The Riemannian metric is either locally constant or can be approximated as such, and \textit{iii)} The manifold is locally connected.
Accordingly, the manifold is modeled with a fuzzy topological structure, and the mapping is found by searching for a low-dimensional projection of the data with the closest possible equivalent fuzzy topological structure.

This Umap approach  was used for creating embedding spaces. The final transformation is given by the function composition $\varphi = ({\cal U} \circ {\cal P})$, where ${\cal U}$ is the nonlinear Umap mapping and ${\cal P}$ is a principal components transformation applied to the nonlinear space with a number of components equal to the number of target dimensions. As a result, $\varphi$ is a canonical mapping where the axes are ordered according to a monotonic decrease of variance in the nonlinear space and will be referred to as the canonical Umap.

\subsubsection{Tree-SNE hierarchical clustering}
\label{mlmethods:Tree-SNE}
Tree-SNE \cite{Robinson-and-Pierce-Hoffman-2020-tree-SNE} is a hierarchical clustering and visualization algorithm based on stacked one-dimensional t-SNE embeddings \cite{van-der-Maaten-and-Hinton-2008}
  with varying perplexity and degrees of freedom of the t-distribution underlying the mapping. As the perplexity and degrees of freedom decrease, smaller and less coarse clusters are produced in the embedding. The approach uses one-dimensional t-SNE embeddings for computational efficiency and ease of clustering via a Fourier transform-based fast \mbox{t-SNE} algorithm \cite{linderman-et-el-2019-fast-t-SNE}, \cite{kobak-et-al-2019-Heavy-tailed-kernels-reveal-a-finer-cluster-structure-in-t-SNE-visualisations}. To that end, a general t-distribution kernel is used ($k(d)= 1/(1+d^2/\alpha)^{\alpha}$, where $d$ is the Euclidean distance between two objects and a scaling parameter $\alpha$ for the degree of freedom $\nu$ given by $\nu = 2\alpha -1$). This reformulation allows kernels with tails heavier than any possible t-distribution to have negative degrees of freedom, which have proven to enhance cluster separability.
  Typically, about 30 to 100 one-dimensional t-SNE embeddings are stacked to create the tree-SNE visualization, with the first layer at the bottom of the plot, and with $\alpha$ and perplexity decreasing for each upper layer. In the tree visualization created using t-SNE, each layer is initialized with the embedding from the previous layer rather than with random initialization. This allows each layer to refine the clustering found in the previous layer, with larger clusters being broken into smaller clusters on subsequent levels of the tree. As a result, the path of a single observation or group of observations can be traced vertically through the tree.

Alpha-clustering is a method that assigns cluster labels to the t-SNE embeddings obtained with Tree-SNE. It recommends the optimal cluster assignment by finding the cluster stability across multiple scales, without prior knowledge of the number of clusters. The group assignment is derived from the t-SNE embeddings obtained with Tree-SNE by automatically selecting the best level of granularity for the data, defined as the stable clustering for which the range of $\alpha$ values is the largest. 
It uses spectral clustering \cite{ng-et-al-2001-On-spectral-clustering_analysis-and-an-algorithm} on each layer of the tree to determine the number of clusters present in the data, rather than using a specified number of clusters. 

The shared nearest neighbors method is used to build the graph for spectral clustering, connecting points in the t-SNE embedding that are nearest neighbors in the original high-dimensional space. Alpha-clustering allows for explainable clustering on multiple levels of granularity. It can be used with tree-SNE visualizations to compare the results to other unsupervised clustering algorithms and determine the meaningful aspects of the data captured by tree-SNE. Alpha-clustering extends and validates the tree-SNE algorithm and provides an opportunity for explainable unsupervised clustering, which is difficult to achieve with other approaches that often rely on deep neural networks. Alpha-clustering helps prevent over-fitting to noise in the data, as it tends to select for clusterings that appear earlier in the tree structure. One-dimensional t-SNE tends to generate easily distinguishable clumps of data, making it easy to identify and count the clumps on each layer of the tree-SNE tree. The difference in magnitude between alpha values rapidly decreases as more layers are added. 

Hierarchical clustering methods have strong limitations posed by the size of the data, but the advantage of the tree-SNE approach is twofold: On the one hand, it has the capability of processing large datasets (prohibitive for classical hierarchical clustering approaches, but required for QM9). On the other hand, the representation mechanism of tree-SNE allows the visualization of dendrograms composed of many thousands of objects, which is the case of QM7b and particularly QM9.

\subsection{Outlier Detection}
\label{mlmethods:Outlier Detection}

Outliers in a dataset are considered rare or odd observations, events, or individuals characterized by unusual attribute values, which deviate from the general distribution observed within a population. Due to their peculiar composition and the fact that they deviate from the main distribution characterizing the data, outliers are also referred to as anomalies.
This definition is inherently subjective and that is why there are several criteria to decide what should be considered abnormal. Outliers may also represent novel or special elements containing valuable information, not necessarily associated with faulty situations. Before considering the possible elimination of these points from the data, it is necessary to understand why they appeared and whether it is likely that similar values will appear again.

Outliers may affect the results of statistical and machine learning procedures, and that is the reason why preprocessing stages use to remove them from the data prior to modeling. However, in the present study outliers are considered a source of information about the molecules. Considering the great care with which QM7b and QM9 were prepared and curated, it is unlikely that outliers would be wrong data that should be disregarded.
Most of the existing outlier detection (OD) algorithms are unsupervised due to the high cost of acquiring ground truth \cite{Zhao-et-al-2019-LSCP-locally-selective-combination-in-parallel-outlier-ensembles} and a good strategy is to use heterogeneous OD approaches, where ensemble methods select and combine diversified base models and produce more reliable results. 
 
The recently introduced SUOD framework \cite{Zhao-et-al-2021-SUOD} enables the combined application of different methods with an ensemble approach based on three independent and complementary levels: \textit{i):} the data level random projection, \textit{ii):} the model level using pseudo-supervised approximations and \textit{iii):} the execution level where the algorithms are applied via balanced parallel scheduling. The data level relies on the Johnson-Lindenstrauss projection \cite{Johnson-and-Lindenstrauss-1984} to create a space for training the models. The model level constructs fast supervised models which use
the unsupervised models’ outputs as “pseudo ground truth”, and the execution level uses a balanced
parallel scheduling mechanism that can forecast the time cost of each OD model (training time). That is done before scheduling so that the task load could be evenly distributed among workers. 

OD was performed with the SUOD approach, using the following combination of state-of-the-art individual techniques: CBLOF 
\cite{He-et-al-2003-Discovering-cluster-based-local-outliers-CBLOF}, HBOS 
 \cite{Goldstein-and-Dengel-2012-HBOS}, LODA \cite{Pevny-2016-Loda-lightweight-on-line-detector-of-anomalies}, LOF \cite{Breunig-et-al-2000-LOF}, and Isolation Forest \cite{Liu-et-al-2008-Isolation-Forest}.

\section{Results}
\label{Results}
\subsection{Unsupervised Analysis}
\label{results:Unsupervised}
The sorted intrinsic dimension estimates obtained with the methods from Section~\ref{mlmethods:Intrinsic-Dimension} are shown in Table~\ref{tab:Intrinsic-Dimension-estimates}, 
\begin{table}[!ht]
\caption{Intrinsic Dimension estimates (sorted).}
\label{tab:Intrinsic-Dimension-estimates}
\centering
\begin{tabular}{|c|c|c||c|c|c|}
\hline
\multicolumn{3}{|c||}{QM7} & \multicolumn{3}{c|}{QM9} \\ \hline
 Method &  IDim      & Params       & Method & IDim & Params             \\ \hline \hline
KNN      & 2.00       &              & DANCo    & 3.07 & k=100              \\ \hline
CorrInt  & 3.08       &              & CorrInt  & 3.41 & k1, k2             \\ \hline
lPCA     & 3.13       & fan,  n7     & lPCA     & 3.68 & fan,  n9           \\ \hline
MiND\_ML  & 3.16      &       n7     & FisherS  & 4.04 &                    \\ \hline
MiND\_ML  & 3.25      &       n7     & MiND\_ML & 4.26 &       n9           \\ \hline
TwoNN    & 3.99       &       n7     & MiND\_ML & 4.30 & mli,  n9           \\ \hline
FisherS  & 4.08       &              & MLE      & 4.36 & mle,  n9           \\ \hline
MLE      & 4.11       & mle,  n7     & TwoNN    & 4.47 &       n9           \\ \hline
TwoNN    & 4.15       &              & TwoNN    & 4.47 &                    \\ \hline
TLE      & 4.90       &       n7     & TLE      & 4.75 &       n9           \\ \hline
MADA     & 5.09       &       n7     & MOM      & 4.88 &       n9           \\ \hline
MOM      & 5.16       &       n7     & FisherS  & 5.10 &       n9           \\ \hline
DANCo    & 5.31       & k=10, n7     & DANCo    & 5.59 & k=10, n9           \\ \hline
FisherS  & 5.48       &       n7     & lPCA     & 7.03 & fo  , n9           \\ \hline
lPCA     & 6.19       & fo,   n7     & DANCo    & 9.00 & k=10               \\ \hline
DANCo    & 6.98       &              & DANCo    & 9.00 &                    \\ \hline
\end{tabular}
\end{table}
where $n7 = 20$, $n9 = 100$ are neighbourhood sizes, $fan, (mli, mle), fo$ are the Fan, maximum likelihood, and Fukunaga-Olsen variants of the corresponding algorithms, respectively. $k1=20$, $k2=100$ (first and second neighborhood sizes considered). The mean and median dimensions are 5.088
and 4.469, respectively, for a $Idim/N_v$ ratio in the $[0.235, 0.268]$ range. Accordingly, most of the information is contained in a subspace of dimension $\approx 25\%$ of the descriptor space.

\subsubsection{Manifold Learning}
\label{results:Manifold Learning}
For visualization purposes and considering the limitations imposed by hard media, 2D approximations of the manifolds containing the data in QM7b and QM9, respectively, can be obtained by computing the mapping $\theta: \mathbb{R}^{19} \rightarrow \mathbb{R}^2$ using Umap (Section~\ref{mlmethods:Manifold-Learning}). They are shown in Fig.~\ref{fig:qm7b-and-qm9-umap-2d-spaces}.
\begin{figure}[!htbp]
\centering
\begin{subfigure}[b]{0.98\columnwidth}
\includegraphics[width=0.95\columnwidth]{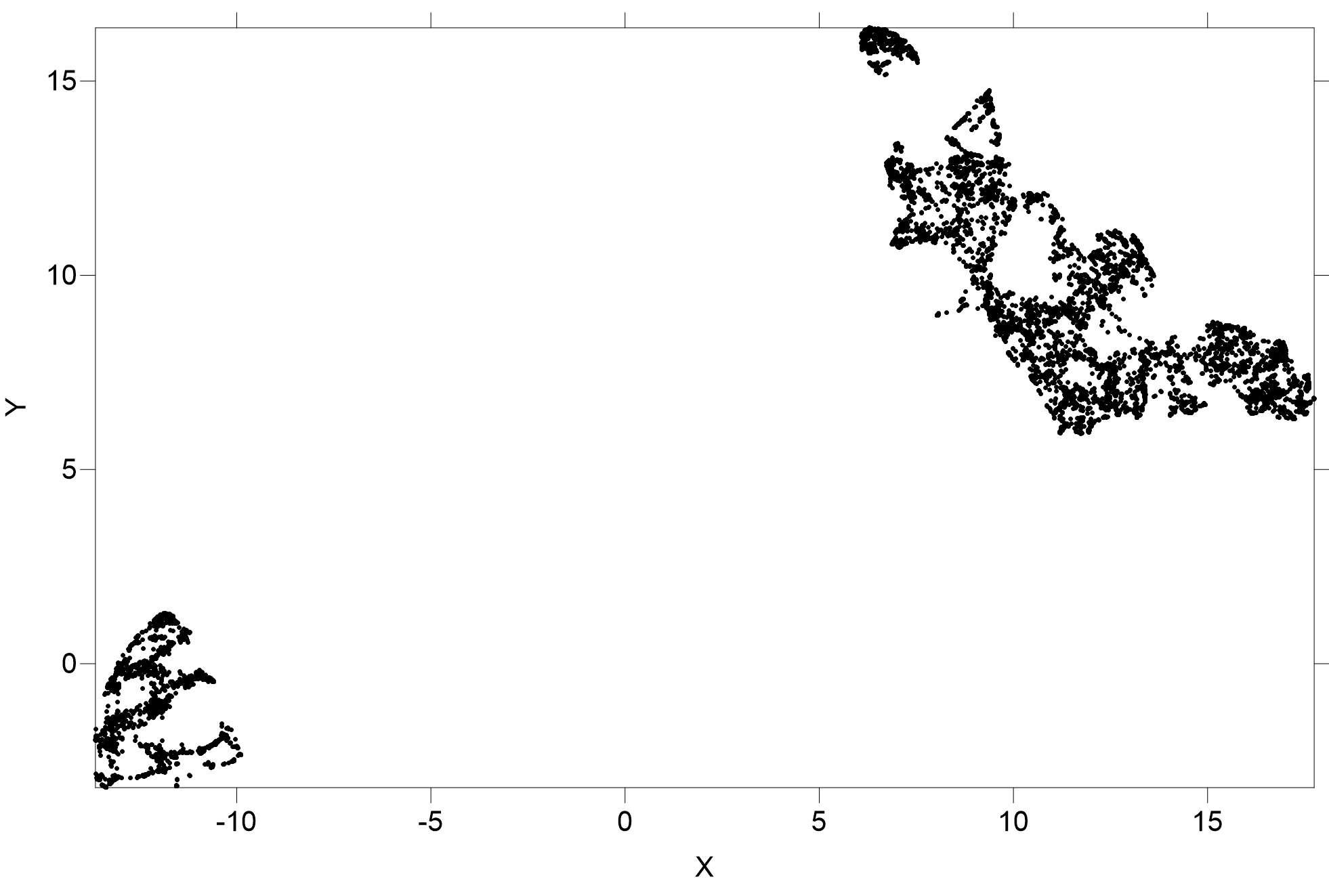}
\caption{QM7b: 7211 molecules}
\label{fig:qm7b-Umap-2d-space}
\end{subfigure} 
\begin{subfigure}[b]{0.9\columnwidth}
\centering
\includegraphics[width=0.8\columnwidth]{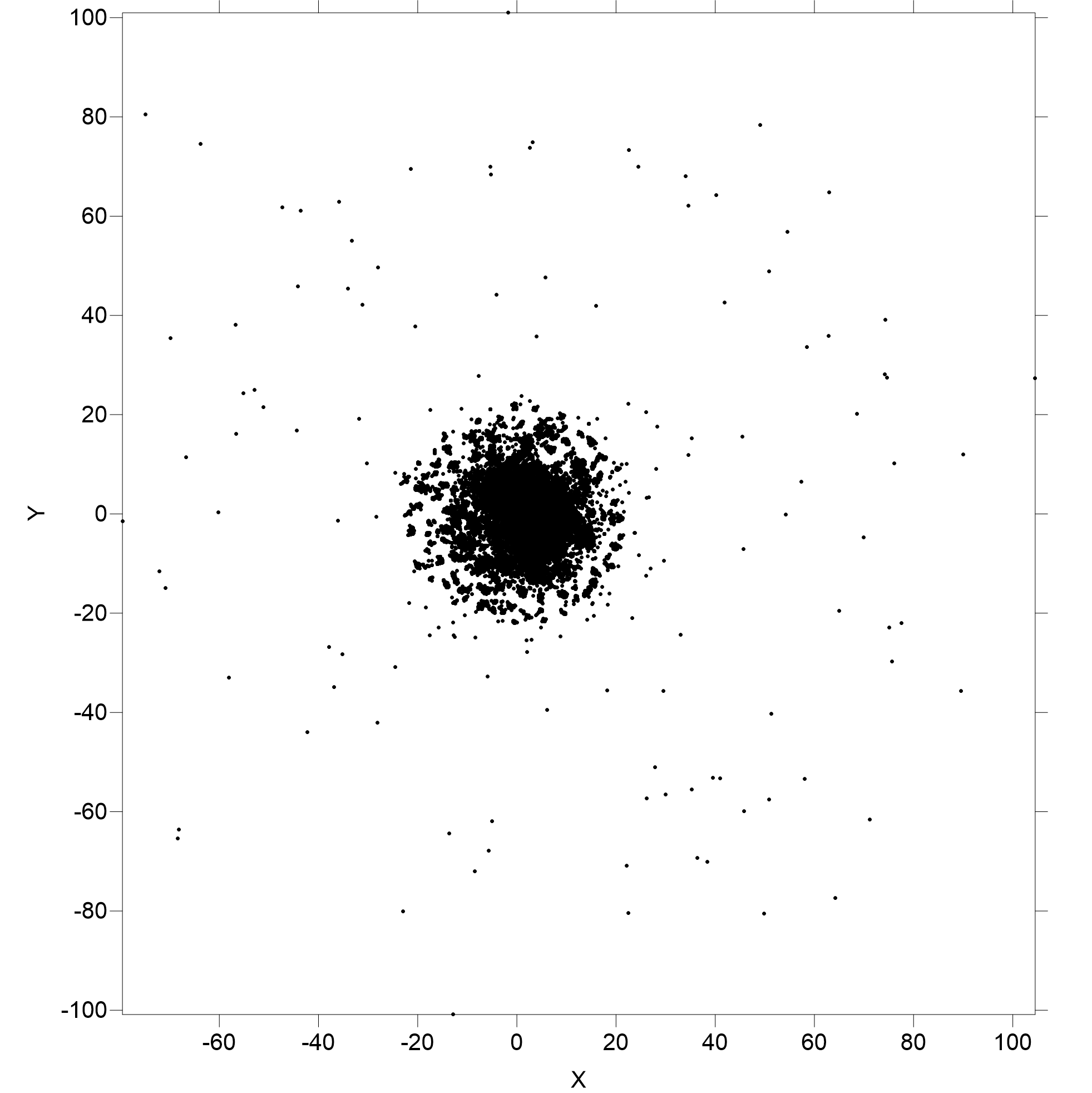}
\caption{QM9: 133885 molecules}
\label{fig:qm9-Umap-2d-space}
\end{subfigure} 
\caption{2D spaces obtained with the Umap mapping.}
\label{fig:qm7b-and-qm9-umap-2d-spaces}
\end{figure}
In the case of QM7b the most distinctive feature is the sharp division of the data into two well-defined clusters of different sizes, with a very large distance in between. The second, larger one exhibits a marked elongation, indicating a strong covariance structure in the nonlinear Umap space. For QM9 the space has a quite different structure, composed of an outer broad region composed of scattered, outlying objects, surrounding an inner much smaller and densely packed core area, concentrating most objects. Both substructures are quasi circular in shape, indicating the lack of covariance between the two embedding space variables. In any case, the observed structures of the 2D embedding spaces are only approximations of the real situation since the intrinsic dimension for both datasets is around five. 

Clustering contributes with another perspective since it was performed in the original descriptor spaces (the spaces defined by the original properties). A hierarchical clustering approach was preferred here in order to avoid methods that require, for example, an estimation of the number of clusters present. Other methods that are free from this restriction, like density-based techniques, require several input parameters, which affect the way in which objects are considered as clusters or isolated elements  (noise of outliers). The aforementioned characteristic of QM9 related to the presence of substructures predominantly composed of outlier elements favor the use of methods that, like hierarchical clustering, do not work with parameters related to the nature of the individuals.

The tree-SNE dendrogram was computed with $100$ alpha levels in order to better appreciate the evolution of the successive embeddings in the hierarchy. The perplexity for t-SNE was automatically started from/reduced to,  a relatively higher value ($365$, because of the large size of the molecule dataset), to a small value ($1.06$). Accordingly, it covered a broad range of coarse/fine substructures.  The number of iterations for the computation of the $\mathbb{R}^{19} \rightarrow \mathbb{R}$ t-SNE mappings required by each level was set to $1000$ to ensure good convergence. The hierarchical trees for QM7b and QM9 are shown in Fig~\ref{fig:qm7-and-qm9-tree-SNE}.
\begin{figure}[!htbp]
\begin{center}
\begin{tabular}{l}
  \begin{subfigure}[b]{0.9\columnwidth}
   \includegraphics[width=0.98\columnwidth, height=0.8\columnwidth]{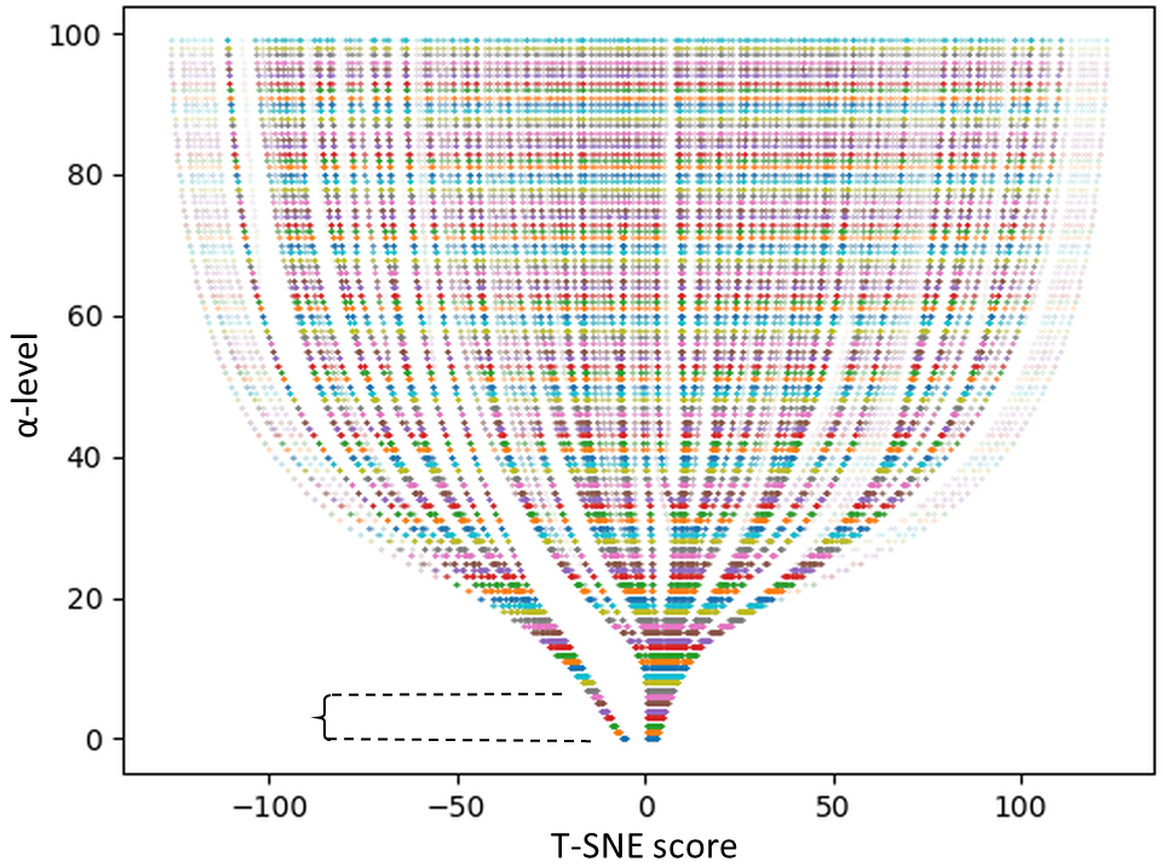}
   \caption{QM7b: The largest cluster stability happens between levels 1 and 6 (the curly-braced region) with 2 clusters .}
   \label{fig:qm7b-tree-SNE}
 \end{subfigure} 
 \\ \\
  \begin{subfigure}[b]{0.9\columnwidth}
   \includegraphics[width=0.98\columnwidth, height=0.8\columnwidth]{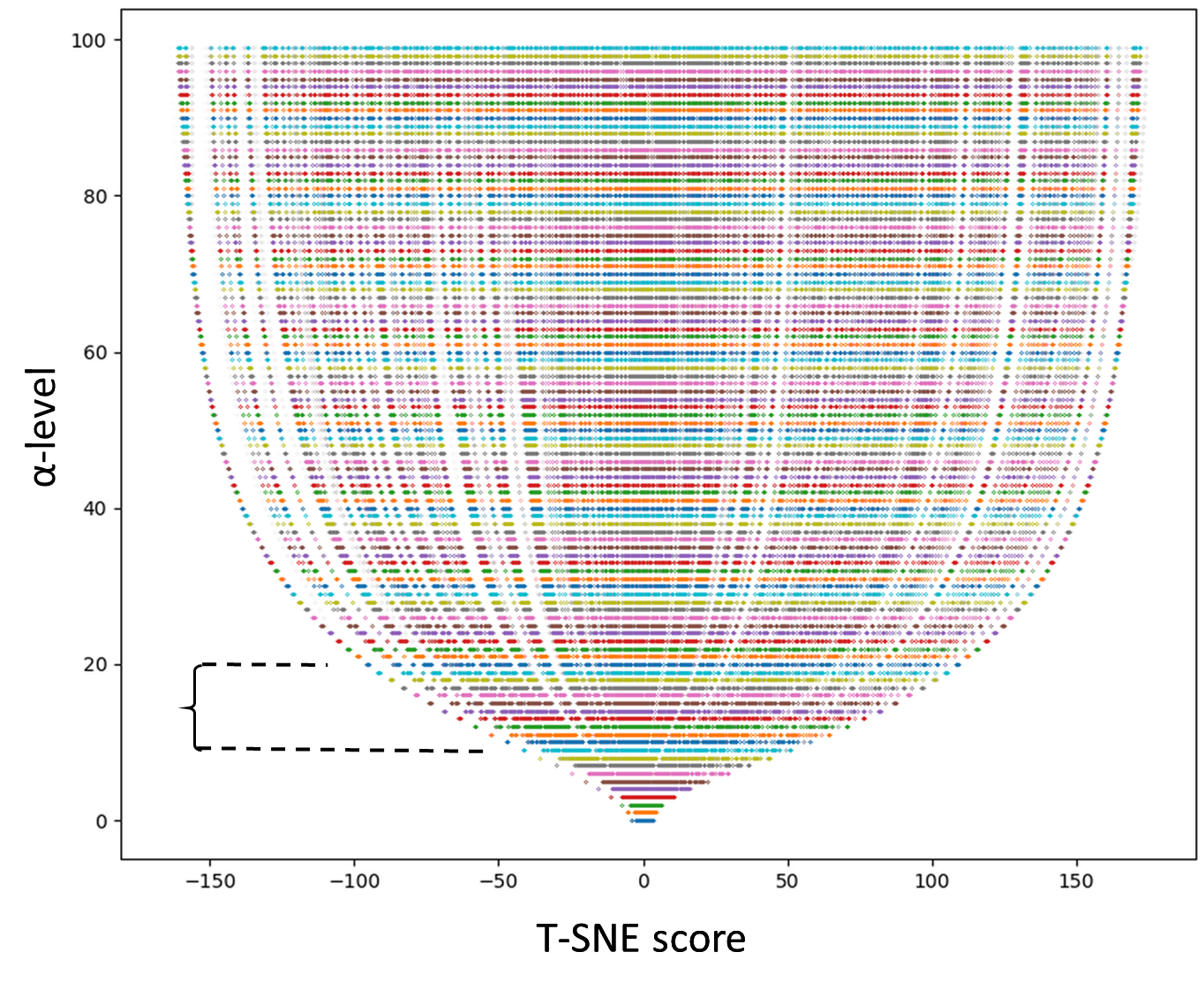}
   \caption{QM9: The largest cluster stability happens between levels 13 and 20 (the curly-braced region) with 33 clusters.}
   \label{fig:qm9-tree-SNE}
 \end{subfigure}
\end{tabular}
\caption{tree-SNE dendrograms with 100 stacked t-SNE levels for the QM7b and QM9 datasets.}
\label{fig:qm7-and-qm9-tree-SNE}
\end{center}
\end{figure}

For QM7b (Fig.~\ref{fig:qm7b-tree-SNE}) the tree exhibits a sharp, distinctive division from the very beginning first $\alpha$-level (1), revealing the existence of two well-differentiated clusters. The widths of the stems at that level provide an indication of cluster size, clearly exposing one smaller and one larger cluster. That differentiation remains clear as the clusters progressively split up to $\alpha$-levels around 25. The curly brace between the horizontal dashed lines indicates the largest $\alpha$-level interval ([1,6]) with a stable number of clusters (2). These results are in good agreement with the structure suggested by the 2D Umap space of Fig.~\ref{fig:qm7b-tree-SNE}.

For QM9 (Fig.~\ref{fig:qm9-Umap-2d-space}) there is no clear division at low $\alpha$-levels and cluster stability is found at higher levels ([13,20]), with 33 clusters, indicated by the curly braces between the horizontal dashed lines. 
 These 33 clusters successively split into smaller ones as perplexity decreases in the hierarchical process, and represents the main constituent of the core region. The cluster separation process in the [13,20] $\alpha$-level interval is visible as gaps that segment the individual lines in Fig~\ref{fig:qm9-tree-SNE}. The structure is very different than that of QM7b, as there are many more clusters, and they are defined at higher $\alpha$ stages. Altogether, the $\alpha$-clustering results (involving all properties) coincide with those found by the exam of the two generated features embedding space of Fig.~\ref{fig:qm9-Umap-2d-space}. 
The interplay between outliers and clusters in QM9 is revealed by the SUOD approach that combines CBLOF, HBOS, LODA, LOF, and Isolation Forest (Section ~\ref{mlmethods:Outlier Detection})  used for characterizing outliers. Based on the empirical probability distribution of the combined SUOD outlier score, a threshold of $0.77$ was set using the $1.5* Interquartile Range$ rule, so that molecules with scores above/under that threshold were classified as outlier/inliner respectively. Accordingly, the number of outliers was $N_{out}=6430$ for an outlier/inliner ratio $N_{out}/N = 4.8\%$.
The combination of the Umap 2D mapping and the outlier analysis is shown in Fig.~\ref{fig:qm9-overall-Umap-2d-mapping-with-inliners-and-outliers}
\begin{figure}[!htbp]
\centering
\begin{tabular}{l}
  \begin{subfigure}[b]{0.9\columnwidth}
     \centerline{\includegraphics[width=0.8\columnwidth]{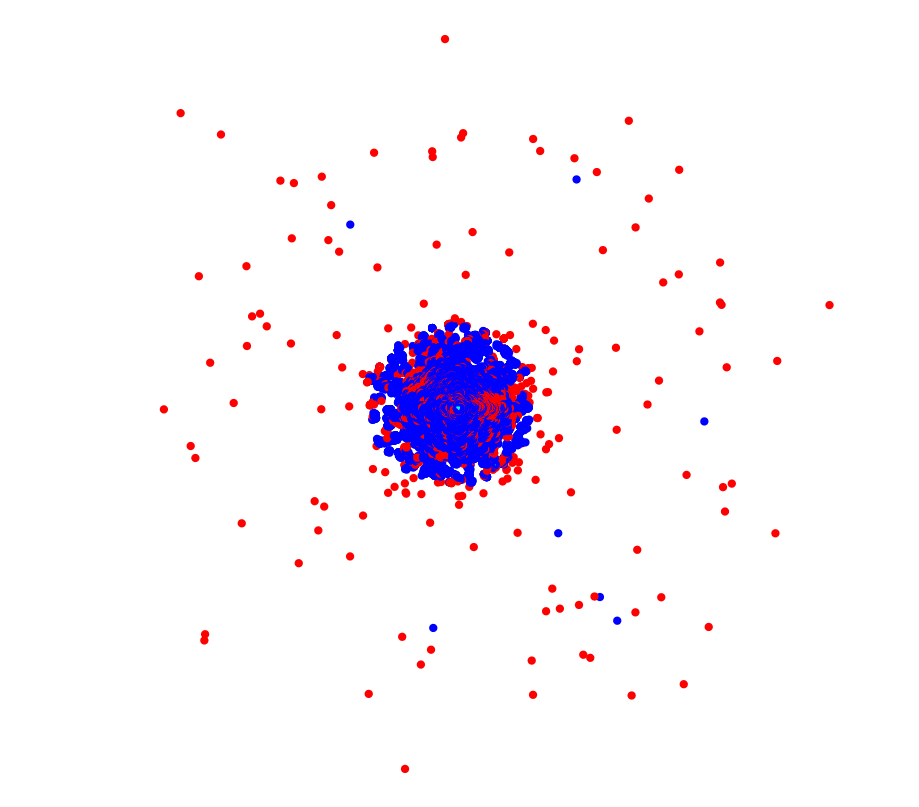}}
     \caption{Overall 2D mapping showing the approximate distribution of inliner (Blue) and outlier (Red) objects.}
     \label{fig:qm9-overall-Umap-2d-mapping-with-inliners-and-outliers}
  \end{subfigure}
\\
  \begin{subfigure}[b]{0.9\columnwidth}
    \centerline{\includegraphics[width=0.8\columnwidth, height=0.7\columnwidth]{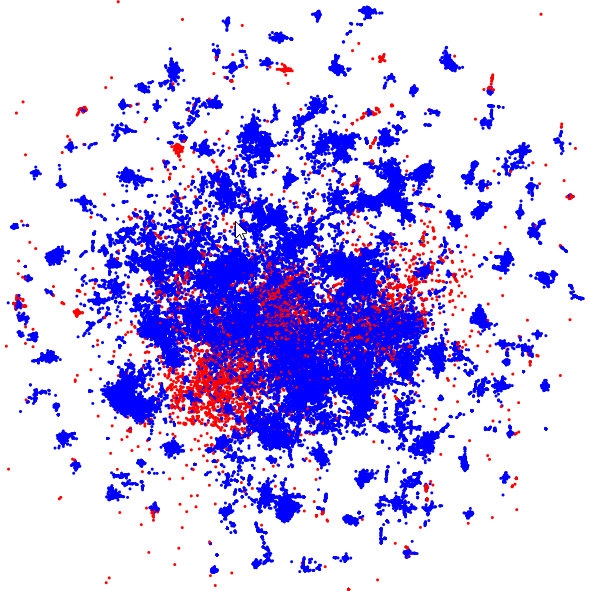}}
     \caption{Zoomed view of the spherical core region of Fig.~\ref{fig:qm9-Umap-2d-space}.}
     \label{fig:qm9-Umap-2D-core-region-closeup}
  \end{subfigure}
\end{tabular}
\caption{Umap 2D space with the 133885 QM9 dataset molecules. Blue: Inliner objects. Red: Outliers.}
\label{fig:qm9-pptk-Umap-2d-space-w-inliners-and-outliers}
\end{figure}
, where the outlier/inliner information is overlapped \textit{a posteriori} to the unsupervised mapping for comparison. Clearly, the outer region is composed predominantly of outliers, whereas the core region concentrates inliner objects in a profusely clustered pattern, previously revealed by tree-SNE clustering. Since the estimated Idim is around 5 (larger that 2), there are inevitable deformations and misplacements of objects in the space. The overall structure is that of a ring-shaped cloud of scattered, outlying points, surrounding a densely packed, much smaller core where most objects are concentrated and clustered. A closer view of the core region exhibits the presence of a large quantity of relatively small to medium-sized clusters (Fig.~\ref{fig:qm9-Umap-2D-core-region-closeup}), with some outliers. 

The dependency between the number of atoms in the
molecule and the outlier score is shown in Fig.~\ref{fig:qm9-outliers-vs-nbrOfAtoms}
\begin{figure}[!htbp]
\begin{center}
\includegraphics[width=1\columnwidth]{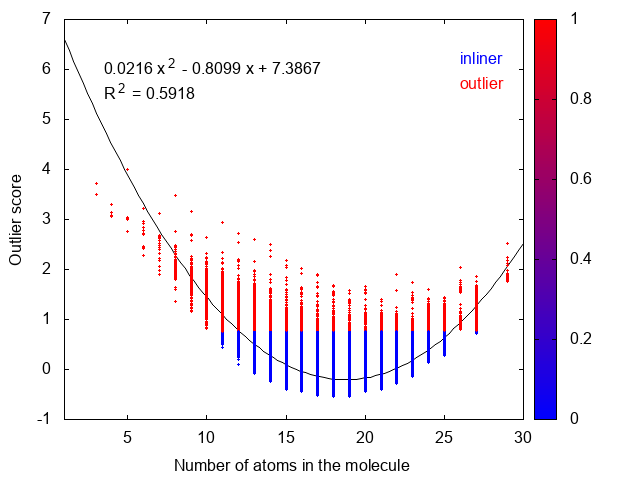}
\caption{Relationship between the number of atoms in the molecule and the outlier score. Blue: Inliners. Red: Outliers. The 2nd degree polynomial regression is least square and $95.2\%$ of the N=133885 points are inliners, determining the location of the middle x-axis range minimum.}
\label{fig:qm9-outliers-vs-nbrOfAtoms}
\end{center}
\end{figure}
, which includes a 2nd order polynomial regression for the outlier scores ($R^2$ is significant at 0.05\%). It reveals that molecules with either small ($< 11$) or large ($> 25$) number of atoms are predominantly outliers, whereas molecules with a number of atoms in the middle range ([11, 25]) are mostly inliners (mainly clustered). Since there is also a relation between the location of the objects in the manifold and the inliner/outlier nature of the molecule, 
 a third dimension was added to the space of Fig.~\ref{fig:qm9-pptk-Umap-2d-space-w-inliners-and-outliers} using as z-axis the number of atoms per molecule ($N_{atoms}$) (a crude indicator of molecule size, determined by molecule composition). Kriging \cite{Matheron-1963-Principles-of-geostatistics}, \cite{Banerjee-et-al-2004-Hierarchical-Modeling-and-Analysis-for-Spatial-Data} was used to create the function $N_{atoms} = f(x,y)$, where $x,y$ are the coordinates of each molecule according to the $\theta$-mapping (kriging produces the best linear unbiased prediction at unsampled locations). The same procedure was applied to QM7b as well and the resulting functions are shown in Fig.~\ref{fig:qm7-umap-2d-space-with-NbrOfAtomsPerMolecule}.
\begin{figure}[!htbp]
\begin{center}
\begin{tabular}{l}
\includegraphics[width=0.98\columnwidth]{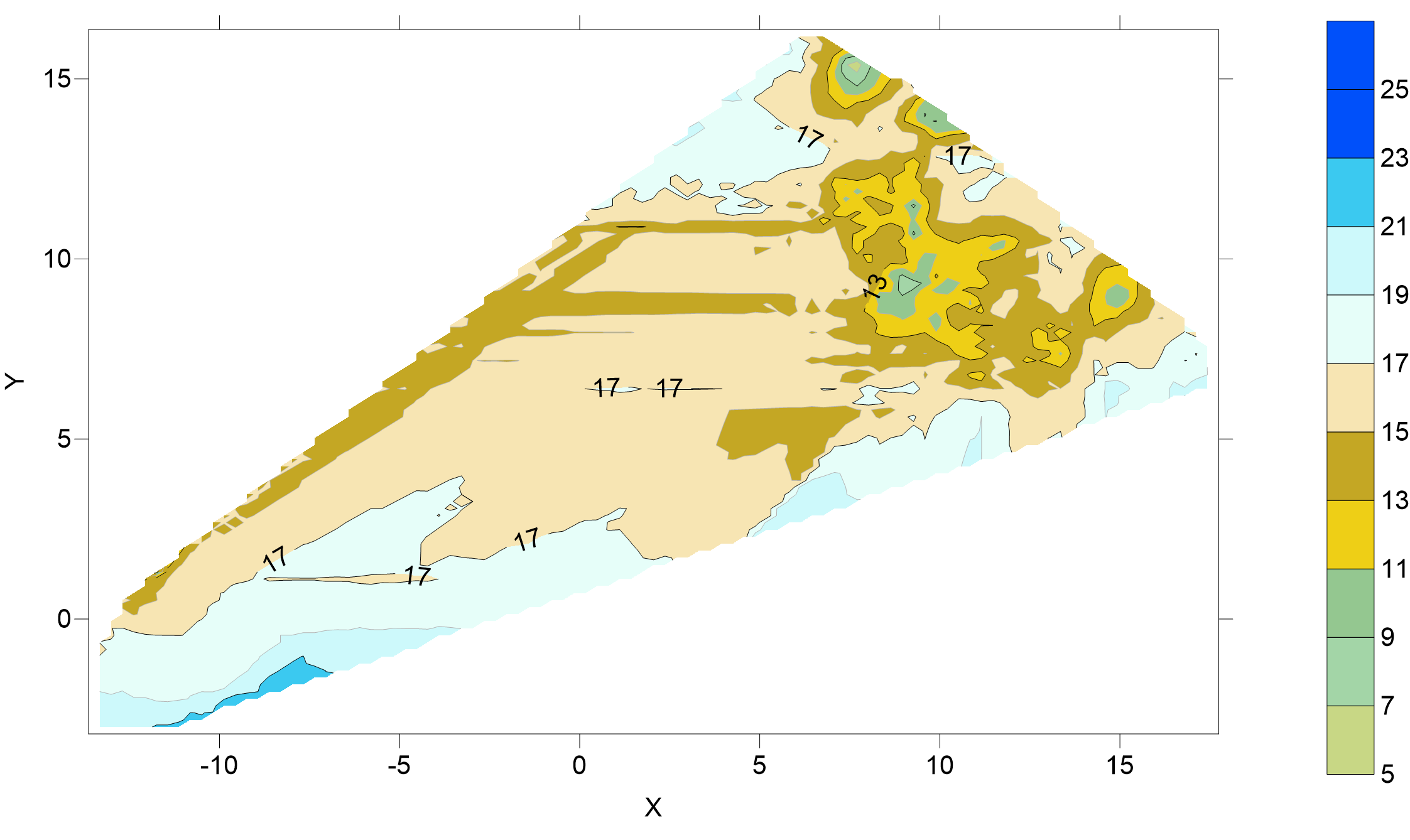}
\\
\includegraphics[width=0.98\columnwidth]{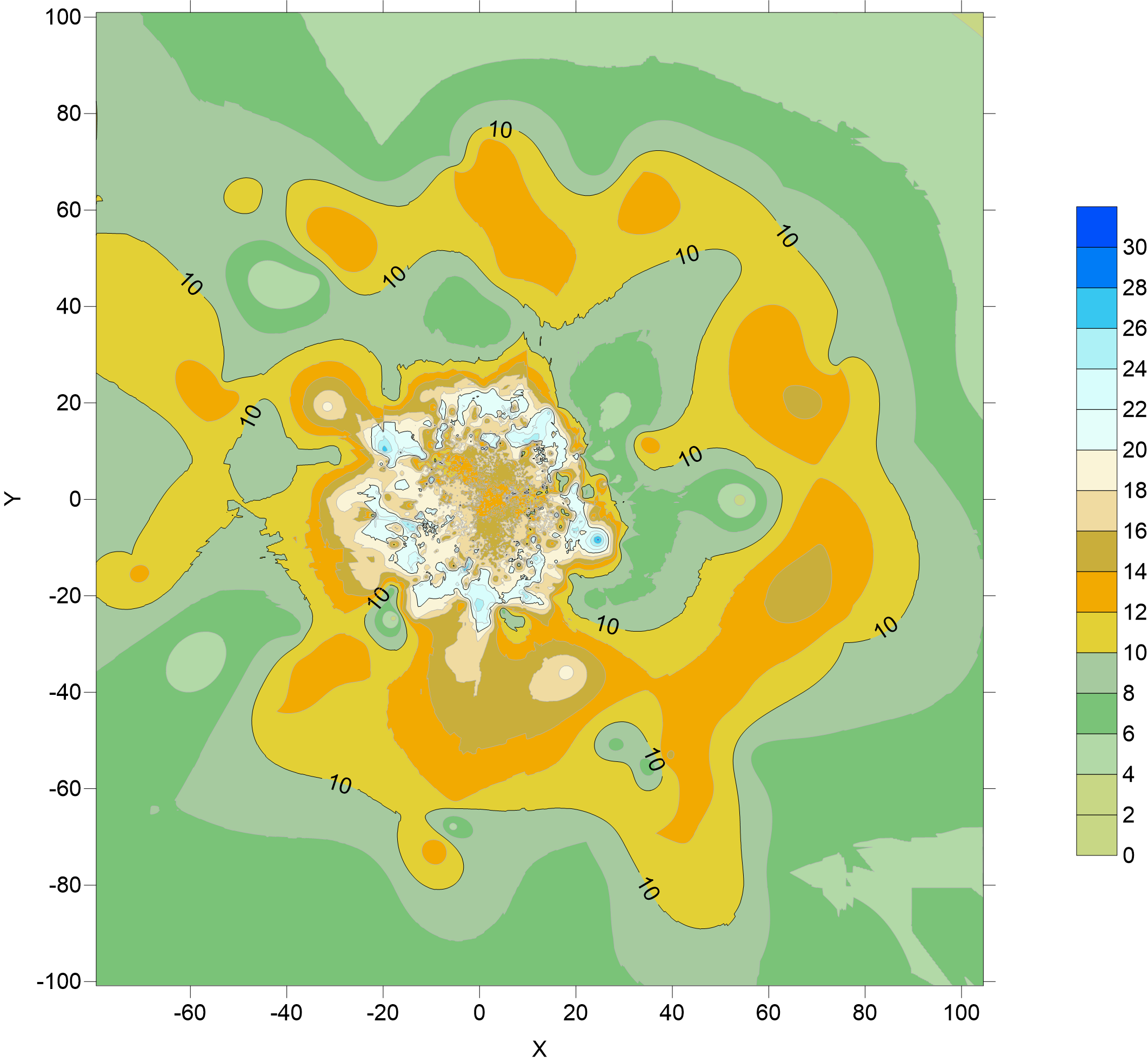}
\end{tabular}
\caption{Distribution of the number of atoms in the molecule in the 2D embedding spaces. Top: QM7. Bottom: QM9.}
\label{fig:qm7-umap-2d-space-with-NbrOfAtomsPerMolecule}
\end{center}
\end{figure}

For QM7b, the lower left corner, which corresponds to the smaller cluster of Fig.~\ref{fig:qm7b-Umap-2d-space}, is predominantly composed of larger molecules ($> 17$ atoms). The upper right region of the space is mostly composed of smaller-size molecules ($< 15$ atoms), with the cluster periphery occupied with larger molecules.
For QM9 there are approximately concentric rings composed of alternating smaller ($< 10$ atoms) and medium ($[10-16]$ atoms) molecules, with an inner core composed of a rim containing large molecules ($> 18$ atoms) encircling a small area of medium-sized molecules. 
A predictability assessment of the functional dependency between the location of the molecules in the 2D manifold and its size (understood as given by the number of atoms in the molecule), was made with regression tree models using a random 90\%/10\% training/testing split. They were used because of a trade-off between their simplicity and their ability to work with either a small or large number of predictor features (the embedding spaces of Fig.~\ref{fig:qm7b-and-qm9-umap-2d-spaces} have only two features, whereas the original descriptor spaces have 14 and 19 respectively). From them, baseline models were produced to illustrate the predictive capability of such spaces, using the number of atoms in the molecule as the target. 
The results for the testing sets of both QM7b and QM9 are shown in Table~\ref{tab:qm7b-and-qm9-m5-rules-for-predicting-the-nbrOfAtoms} and Fig.~\ref{fig:qm7-and-qm9-m5rules}.
\begin{table}[!ht]
    \centering
    \begin{tabular}{|c|c|c|c|c|c|}
     \hline
     \multicolumn{6}{|c|}{QM7b} \\ \hline
     Features       &  MSE    & RMSE   & MAE    & R      & nbrOfRules  \\ \hline \hline
       2  generated & 1.0525 & 1.0259 & 0.7146 & 0.9266  & 70         \\ \hline
       14 original  & 0.1397 & 0.3737 & 0.2451 & 0.9906  & 101         \\ \hline
     \hline
     \multicolumn{6}{|c|}{QM9} \\ \hline
     Features       &  MSE    & RMSE   & MAE    & R      & nbrOfRules  \\ \hline \hline
       2  generated & 0.8917  & 0.9443 & 0.3382 & 0.9475 & 358         \\ \hline
       14 original  & 0.0061  & 0.0781 & 0.0315 & 0.9997 & 46          \\ \hline
    \end{tabular}
    \caption{Regression tree models for QM7b and QM9 predicting the number of atoms in the molecule.}
    \label{tab:qm7b-and-qm9-m5-rules-for-predicting-the-nbrOfAtoms}
\end{table}
%
%
\begin{figure}[!htbp]
\begin{center}
  \begin{subfigure}[b]{\columnwidth}
   \includegraphics[width=0.48\columnwidth]{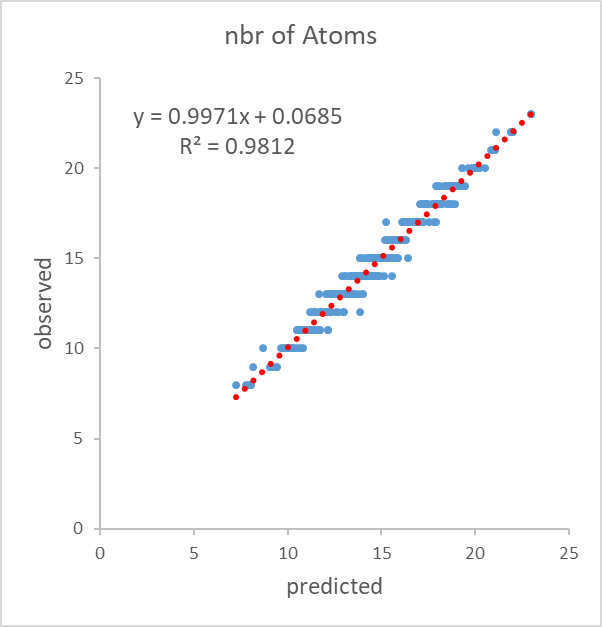}
   \includegraphics[width=0.48\columnwidth]{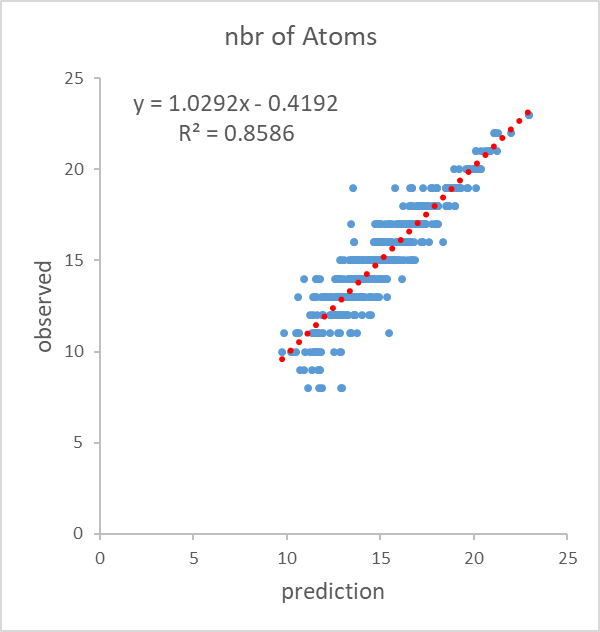}
   \caption{QM7b dataset: Relation between the predicted and observed number of atoms in the molecule for the testing set (Left: model with all features. Right: model with the 2 generated features).}
   \label{fig:qm7-m5rules}
 \end{subfigure}
 \\
  \begin{subfigure}[b]{\columnwidth}
   \includegraphics[width=0.48\columnwidth]{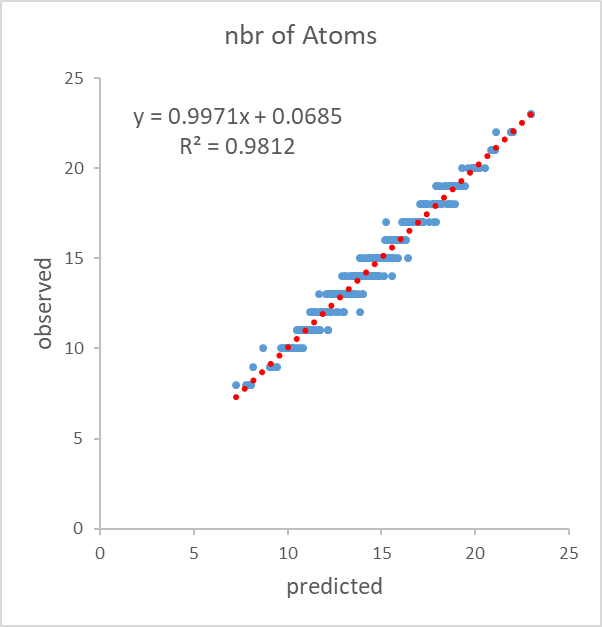}
   \includegraphics[width=0.48\columnwidth]{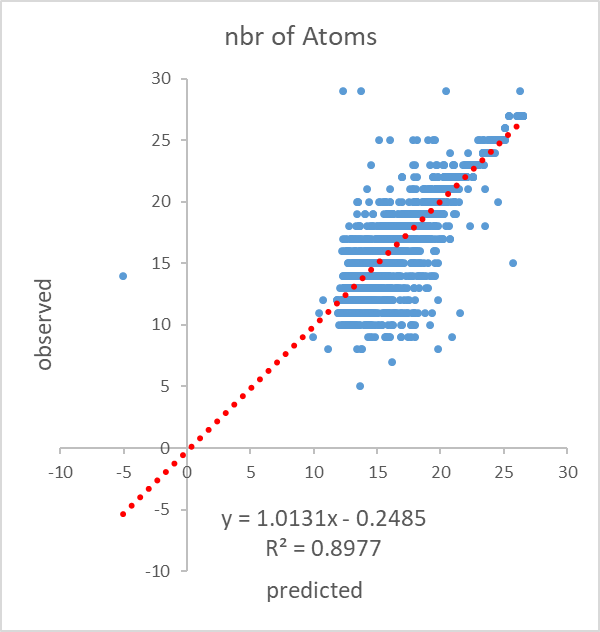}
   \caption{QM9 dataset: Relation between the predicted and observed number of atoms in the molecule for the testing set (Left: model with all features. Right: model with the 2 generated features).}
   \label{fig:qm9-m5rules}
 \end{subfigure}
\caption{Predicted and observed plots for the number of atoms in the molecule (testing sets) for QM7b and QM9.}
\label{fig:qm7-and-qm9-m5rules}
\end{center}
\end{figure}
For the two datasets, highly accurate models were obtained when using the original properties, as evidenced by the low error measures, the high correlations, and the behavior of the prediction vs. the observations. When the predictions are made from small dimension embedding spaces the errors are much higher, but the correlations and only slightly smaller. The predictions vs. the observations are more scattered, particularly for QM9, but still with a highly significant R2. Considering that the dimension of the embedding space is only 2, much smaller than the 
mean of the intrinsic dimension estimate, achieving this level of predictability is noteworthy.

 

\section{Conclusions}
\label{Conclusions}
Unsupervised analysis revealed that the overall data structure of the QM7b and QM9 molecule datasets are quite different. While the former has a clearly defined two clusters structure, the latter is composed of an outer region predominantly containing outliers and an inner, core region that concentrates inliner objects grouped in clusters. For QM9 there is a significant relationship between the number of atoms in the molecule, and its outlier/inliner nature so that molecules with an either small or large number of atoms are predominantly outliers, whereas molecules with number of atoms in the middle range are mostly inliners and clustered. These characteristics should be taken into account when developing predictive models for inverse-design of \textit{de-novo} molecules.

For both datasets the intrinsic dimension is several times smaller than the descriptor dimension, indicating a high level of redundancy. Despite the structural differences, their properties carry strong predictive information for molecular composition, as illustrated by the number of atoms in the molecule, which can be accurately predicted from the original properties. Embedding spaces of small dimensions still retain predictive capabilities.

%
%





\end{document}